\documentclass[aps,prd,preprintnumbers,superscriptaddress,nofootinbib,floatfix,twocolumn,notitlepage]{revtex4-1}
\usepackage{graphicx}  
\usepackage{dcolumn}   
\usepackage{bm}        
\usepackage{epsfig,amsmath,amssymb,verbatim,mathrsfs,array,layout,textcomp,amssymb,latexsym,slashed}
\usepackage{xcolor}
\usepackage[colorlinks=true,citecolor=blue,urlcolor=blue,linktocpage=true,
linkcolor=blue]{hyperref}
\usepackage{cleveref}
\usepackage[utf8]{inputenc}
\usepackage{multirow}
\usepackage{aas_macros}

\usepackage[compat=1.0.0]{tikz-feynman}
\usetikzlibrary{arrows,shapes}
\usetikzlibrary{trees}
\usetikzlibrary{matrix} 
\usetikzlibrary{positioning}				
\usetikzlibrary{calc,through}				
\usetikzlibrary{decorations.pathreplacing}  
\usepackage{pgffor}							
\usetikzlibrary{decorations.pathmorphing}	
\usetikzlibrary{decorations.markings}

\newcommand{\cO}{\mathcal{O}}

\def\beq{\begin{equation}}
\def\eeq{\end{equation}}
\def\beqa{\begin{eqnarray}}
\def\eeqa{\end{eqnarray}}

\begin{document}

\title{Rekindling s-Wave Dark Matter Annihilation Below 10$\,$GeV with Breit-Wigner Effects
}
\preprint{IPPP/25/14\,,\ LAPTH-009/25}


%
\author{Genevi\`eve B\'elanger}
\email{genevieve.belanger@lapth.cnrs.fr}
\affiliation{Laboratoire d'Annecy de Physique Th\'eorique, CNRS -- USMB, 74940 Annecy, France}
\author{Sreemanti Chakraborti}
\email{sreemanti.chakraborti@durham.ac.uk}
\affiliation{Institute for Particle Physics Phenomenology, Department of Physics
Durham University, \\ Durham, DH1 3LE, United Kingdom}
\author{C\'edric Delaunay}
\email{cedric.delaunay@lapth.cnrs.fr}
\affiliation{Laboratoire d'Annecy de Physique Th\'eorique, CNRS -- USMB, 74940 Annecy, France}
\author{Margaux Jomain}
\email{margaux.jomain@lapth.cnrs.fr}
\affiliation{Laboratoire d'Annecy de Physique Th\'eorique, CNRS -- USMB, 74940 Annecy, France}
%


\begin{abstract}
Velocity-independent (s-wave) annihilation of thermal Dark Matter is ruled out by CMB data for masses below $\sim10\,$GeV, effectively ruling out the possibility of indirectly detecting it in this mass range. We demonstrate in a model-independent framework that Breit-Wigner effects from very narrow resonances can circumvent CMB constraints, thereby reviving the potential to detect s-wave DM annihilation in the present Universe. The density of resonant s-wave Dark Matter continues to evolve long after chemical decoupling, leading to a scenario we refer to as {\it belated freeze-out}, where kinetic decoupling plays a significant role in determining the relic density. 
\end{abstract}

\maketitle

\section{Introduction}

We see the cosmos bright, yet we know it is painted black. The mystery persists: where does dark matter (DM) hide its track?  
Thermal freeze-out~\cite{Lee:1977ua,Srednicki:1988ce,Gondolo:1990dk} remains one of the most compelling mechanisms for producing the DM relic density in the early Universe. However, minimal DM models face strong constraints from a wide range of experimental searches. 
In particular, weakly interacting DM particles below the electroweak scale are tightly constrained by direct and indirect detection, cosmology, and collider experiments (see {\it e.g.}~\cite{Cirelli:2024ssz} for a recent review).\\

Moreover, since these constraints are governed by DM interactions with the Standard Model (SM), they are often strongly correlated. 
This correlation can be alleviated if DM annihilation is velocity-dependent. This occurs in p-wave annihilation or even in s-wave when annihilation proceeds via an s-channel resonance. In the latter case, 
 a Breit-Wigner (BW) resonance enhancement at the low DM velocities found in the galactic halo was proposed to explain observed excesses in indirect detection which suggest an annihilation cross-section exceeding the thermal value, $\langle \sigma v\rangle_{\rm th} \approx 3\times 10^{-26}{\rm cm}^3/s$ ~\cite{Ibe:2008ye,Feldman:2008xs,Guo:2009aj}.  
BW enhancement has since been explored as a mechanism for boosting DM annihilation in galaxies in a variety of models~\cite{AlbornozVasquez:2011js,Ding:2021sbj}. 

The same effect has been proposed as a way to evade cosmic microwave background (CMB) constraints on GeV-scale DM. Energy injected into the primordial plasma by electrons and photons from  DM annihilation affects CMB anisotropies, typically excluding thermal DM for masses below $\sim10\,$GeV~\cite{Slatyer:2015jla}. However, by leveraging the velocity dependence of DM annihilation, these constraints can be avoided (since the DM velocity is very small  during the recombination era) while still allowing sufficient annihilation during freeze-out. This mechanism has been explored for p-wave annihilation~\cite{Belanger:2024bro,Chen:2024njd,Wang:2025tdx} and is also relevant for s-wave DM annihilation when it occurs near a resonance~\cite{Bernreuther:2020koj,Binder:2022pmf,Cheng:2023dau,Balan:2024cmq,Wang:2025tdx}. In contrast to the p-wave and higher-order wave proposals for evading CMB constraints, signals of resonant s-wave DM annihilation are not significantly suppressed and remain potentially observable in the present Universe. Another way to evade CMB constraints while retaining strong prospects for indirect detection is through inelastic DM~\cite{Berlin:2023qco}.\\

In this work, we take  a model-independent approach to s-wave resonant annihilation for DM below the electroweak scale. We determine the allowed parameter space consistent with both relic density and CMB constraints, adjusting the mediator's mass and width so that the resonance enhancement occurs primarily at the time of DM formation, thus avoiding CMB limits. In particular, we show for various annihilation channels that reproducing the DM relic density while avoiding CMB limits generically requires very narrow resonances with width-to-mass ratio smaller than $\sim 10^{-7}$, depending on DM and mediator masses. 

A key feature of resonant annihilation is that DM production extends to much lower temperatures than in standard freeze-out~\cite{Ibe:2008ye}.
We highlight that the small couplings typically required to achieve the correct relic density in the presence of a resonance lead to kinetic decoupling of DM during belated freeze-out, significantly impacting DM production~\cite{Binder:2017rgn,Binder:2021bmg,Duch:2017nbe}. To account for this effect, we estimate the decoupling temperature in a model-independent manner and analytically solve the modified Boltzmann equation governing DM number density evolution, assuming dark sector self-interactions are sufficiently rapid for dark states to establish a thermal equilibrium with their own temperature. 

We also investigate the impact of DM indirect detection constraints from various astrophysical observations at different distance scales, including XMM-Newton~\cite{Cirelli:2023tnx} (galactic halo), Fermi-LAT~\cite{Fermi-LAT:2015att} (dwarf galaxies), and MeerKAT~\cite{Lavis:2023jju} (galaxy clusters).  We emphasize how  current constraints extracted from observations involving different DM velocity regimes depend on the resonance parameters. Our results show that thermal  s-wave DM annihilation remains viable below $\sim 10\,$GeV and presents promising targets for future indirect detection searches, provided the dark sector parameters enable a BW enhancement both during DM formation and in the present Universe. 
In our model-independent approach, we  do not discuss direct detection constraints. We expect those to be suppressed  due to the small couplings involved~\cite{Binder:2022pmf,Belanger:2024bro}.\\

The remainder of this paper is organised as follows. \Cref{sec:resan} provides a general discussion of resonant s-wave annihilation, while~\cref{sec:belatedFO} explores its impact on the DM abundance. \Cref{sec:kindec} summarizes key aspects of kinetic decoupling. \Cref{sec:CMBlimits} presents constraints from CMB anisotropies, and~\cref{sec:IDsignals} examines current limits from indirect DM searches and presents predictions from BW enhanced s-wave DM annihilation in various astrophysical contexts. \Cref{sec:conclusions} concludes with a summary of our findings. The appendices include details on the computation of the relic density with and without kinetic decoupling, the average DM velocity after chemical decoupling, and the derivation of CMB bounds.

\section{Resonant s-wave Annihilation}\label{sec:resan}

Consider a DM particle $\chi$ of spin $S_\chi$ and mass $m_\chi$, initially in thermal equilibrium with the SM plasma at early times ($T\gg m_\chi$). Near a resonance, its annihilation cross-section takes a generic BW form,
\beq\label{eq:sigmaBW}
\sigma_{\chi\chi\to f}=\frac{4\pi \omega}{ p^2 }B_\chi B_f\frac{m_R^2 \Gamma_R^2}{(s-m_R^2)^2+m_R^2\Gamma_R^2}\,,
\eeq
with $\omega\equiv (2J_R+1)/(2S_\chi+1)^2$, and $p\equiv \sqrt{s-4m_\chi^2}/2$ is the DM center-of-mass momentum in terms of the total energy $\sqrt{s}$. The parameters $J_R$, $m_R$, and $\Gamma_R$ denote the spin, mass, and total width of the resonance, respectively. The branching ratios are given by $B_\chi$ for resonance decay into DM pairs and $B_f=1-B_\chi$ for other final states.  
Since the (numerator of the) branching ratios in Eq.~\eqref{eq:sigmaBW} are energy-dependent, and assuming s-wave dominance with negligible velocity-suppressed corrections, their product near the resonance follows, 
\beq
B_\chi B_f\simeq \frac{p}{\bar p} \times \bar B_\chi \bar B_f\,,  
\eeq
where barred quantities denote their ``on-shell" values at $\sqrt{s}=m_R$.  
Introducing the dimensionless parameters $\epsilon_R\equiv m_R^2/(4m_\chi^2)-1$ and $\gamma_R\equiv m_R\Gamma_R/(4m_\chi^2)$, the annihilation cross-section times the lab-frame DM velocity reads~\cite{Gondolo:1990dk},
\beq\label{eq:sigmav}
\sigma v_{\rm lab}= \frac{8\pi b_R (1+\epsilon)^{1/2}}{m_\chi^2(1+2\epsilon)\epsilon_R^{1/2} }\left[\frac{\gamma_R^2}{\gamma_R^2+(\epsilon-\epsilon_R)^2}\right]\,,
\eeq
where $\epsilon\equiv (s-4m_\chi^2)/(4m_\chi^2)\simeq v_{\rm lab}^2/4$ is the squared relative DM velocity,  and we define $b_R\equiv \omega \bar B_\chi(1-\bar B_\chi)\leq \omega/4$. \\

To evade strong CMB constraints while ensuring resonant enhancement at freeze-out, we impose $\epsilon_{\rm CMB}\ll \epsilon_R\lesssim 1$. When $\epsilon_R\gg \epsilon_{\rm halo}\sim 10^{-6}$, DM annihilation is only weakly enhanced in the Milky Way, limiting the indirect detection prospects. As we demonstrate below, existing searches already exclude resonant s-wave DM with $\epsilon_R\lesssim 10^{-6}$. We thus focus on $\epsilon_R$ in the range $10^{-4}- 10^{-6}$ in the following. Furthermore, for consistency of the narrow-width approximation we restrict our analysis to values of $\gamma_R\lesssim\epsilon_R$.

Such a fine-tuned mass coincidence between DM and the mediator is not natural in the 't Hooft sense within minimal models. A notable exception arises in Kaluza-Klein DM models~\cite{Kakizaki:2005en} based on large extra dimensions~\cite{Arkani-Hamed:1998jmv,Arkani-Hamed:1998sfv}, where $m_R=2m_\chi$ can be achieved at tree level. Nevertheless, this scenario remains phenomenologically compelling, representing one of the last viable scenarios for thermal s-wave DM  below $m_\chi\sim 10\,$GeV. A key feature of these resonances is their significant impact on freeze-out dynamics, delaying it well beyond the point of chemical decoupling from the thermal bath.

\section{Belated Freeze-out}\label{sec:belatedFO}

We assume that DM particles were initially in chemical equilibrium with the SM in the early Universe and that the present relic density of DM has formed through the freeze-out mechanism. 
The Boltzmann equation governing the number density of DM is given by~\cite{Lee:1977ua,Bernstein:1985th,Scherrer:1985zt},
\beq\label{eq:BEq}
\dot n_\chi +3H n_\chi = -\langle \sigma v\rangle (n_\chi^2-n_{\chi\,\rm eq}^2)\,,
\eeq
where $H$ is the Hubble parameter, $n_{\chi\,\rm eq}$ denotes the  equilibrium DM density, and $\langle \sigma v\rangle$ is the  thermally-averaged annihilation cross-section. 

Following standard freeze-out calculations for resonant annihilation~\cite{Gondolo:1990dk} which are briefly reviewed in~\cref{app:resonantFO}, the contribution of $\chi$ to the energy budget of the Universe today is approximately,
\beq\label{eq:Omchi}
\Omega_\chi h^2\simeq 5.5\times 10^{-13} N_\chi \frac{m_{\chi{\rm GeV}}^2\epsilon_R^{1/2}}{b_R\gamma_R\bar g_\star^{1/2}}\,,
\eeq
where $N_\chi=1$ ($N_\chi=2$) when $\chi$ and $\bar\chi$ are (not) identical particles, $m_{\chi{\rm GeV}}$ is the DM mass in GeV, and  $\bar g_\star^{1/2}$ is the effective number of degree-of-freedom of the SM plasma at the time of chemical decoupling.   
This expression assumes that DM exchanges energy with the SM sufficiently fast to maintain a thermal distribution throughout freeze-out. However, for highly effective resonant annihilation, this kinetic-equilibrium condition is typically not met~\cite{Binder:2017rgn}, as the DM-SM scattering cross-section does not benefit from the resonance and is suppressed by small couplings. In such scenarios, the Boltzmann equation must be solved directly for the DM momentum distribution rather than the number density. This is a challenging numerical task~\cite{Binder:2021bmg}, requiring detailed modeling to relate the scattering and annihilation cross-sections. To keep the discussion general, we parameterize corrections to~\cref{eq:Omchi} from early kinetic decoupling with a model-dependent normalization factor $k_{\rm dec}$. Thus, we write,
\beq\label{eq:relicdensity}
b_R\gamma_R\epsilon_R^{-1/2}=4.6\times 10^{-12}k_{\rm dec}N_\chi \frac{m_{\chi{\rm GeV}}^2}{\bar g_\star^{1/2}f_\chi}\,,
\eeq
where $f_\chi\equiv \Omega_\chi/\Omega_{\rm DM}$ is the fraction of the total amount of DM observed, with $\Omega_{\rm DM}h^2\approx 0.12$, that is in the form of $\chi$ (and eventually $\bar\chi$) particles. We will show in~\cref{sec:kindec} that for $\epsilon_R\ll 1$, $k_{\rm dec}$ is expected to be much smaller than unity, meaning the relic density is strongly suppressed, with the approximate scaling $k_{\rm dec} \propto\sqrt{\epsilon_R}$.\\

\begin{figure}
    \centering
    \includegraphics[width=0.5\textwidth]{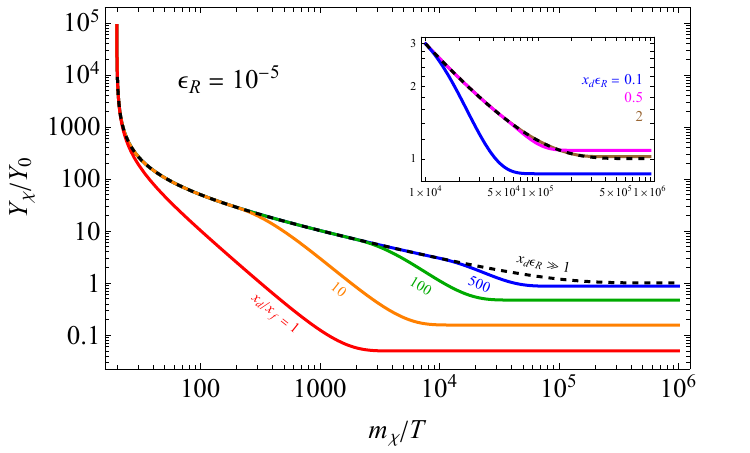}
    \caption{DM abundance $Y_\chi$ as function of time, parameterized by $x= m_\chi/T$, after chemical decoupling at $x_f=20$ in the belated freeze-out scenario for $\epsilon_R=10^{-5}$. The dashed line assumes kinetic equilibrium ($x_d\epsilon_R\gg 1$), while the red, orange, green and blue curves represent  the suppression due to kinetic decoupling at $x_d/x_f=1$, 10, 100, and 500, respectively. The inset highlights the mild  enhancement of the abundance for very late decoupling at $x_d\epsilon_R=1/2$ (magenta) and 2 (brown). Abundances are shown in units of the kinetic equilibrium value at $x\to \infty$, denoted as $Y_0$.}
    \label{fig:yield}
\end{figure}

\Cref{fig:yield} shows that for $\epsilon_R\ll 1$, the abundance of $\chi$ continues to evolve long after chemical decoupling at $x_f\approx 20$, resulting in a belated freeze-out regime. Although the annihilation rate has fallen below the Hubble expansion rate, the annihilation cross-section continues to grow at lower temperatures~\cite{Ibe:2008ye}. As DM particles cool, the peak of their velocity distribution aligns more closely with the resonance peak. Consequently, the DM yield slowly fades out, scaling approximately as $x^{-1/2}$ where $x\equiv m_\chi/T$, until $x\sim \epsilon_R^{-1}$, at which point DM particles are too cold to satisfy the resonance condition. While this description remains qualitatively correct, it is significantly modified by kinetic decoupling, which, as we argue below, likely occurs in the course of this evolution.

\section{Kinetic decoupling}\label{sec:kindec}

Resonant annihilation typically requires much smaller coupling values to achieve the correct DM relic density. This generally leads to kinetic decoupling of DM during the freeze-out process. Below, we estimate the kinetic-decoupling correction parameter,  
\beq
k_{\rm dec}\equiv \frac{\Omega_\chi}{\Omega_\chi^{\rm keq}}\,,
\eeq
where $\Omega_\chi^{\rm keq}$ denotes the abundance obtained from~\cref{eq:Omchi} under the assumption of kinetic equilibrium. 

We assume instantaneous kinetic decoupling and that DM particles are in sufficiently fast self-interactions, allowing them to form a dark equilibrium at temperature $T'\leq T$ after decoupling at $T=T_d$. For $T\geq T_d$, $T'=T$, while for $T<T_d$, the DM temperature evolves with time according to,  
\beq\label{eq:Tprime}
T'(T<T_d)\equiv \xi(T) T\,,\quad \xi(T)=
\left[\frac{h_{\rm eff}(T)}{h_{\rm eff}(T_d)}\right]^{2/3}\frac{T}{T_d}\,.
\eeq
The linear scaling with $T$ follows from the fact that DM is non-relativistic at decoupling, while the prefactor captures the subsequent reheating of the SM plasma due to the decoupling of SM species; $h_{\rm eff}$ denotes the number of degrees of freedom of the entropy density. 

The right-hand side of the Boltzmann equation in~\cref{eq:BEq} is modified to,
\begin{align}
\dot n_\chi +3H n_\chi =& -\langle \sigma v\rangle' n_\chi^2+\langle \sigma v\rangle n_{\chi\,{\rm eq}}^2\nonumber\\
=&-\langle \sigma v\rangle\left[\beta(\xi)n_\chi^2-n_{\chi\,{\rm eq}}^2\right]\,,\label{eq:BEq2}
\end{align}
where $\langle \sigma v\rangle'$ is obtained by taking  $T\rightarrow T'$ in $\langle \sigma v\rangle$, and the second line assumes resonant annihilation, with,
\beq\label{eq:beta}
\beta(\xi)\equiv \xi^{-3/2}e^{-x\epsilon_R(\xi^{-1}-1)}\,.
\eeq
We show in~\cref{app:kdec} that solving~\cref{eq:BEq2} approximately yields,
\beq\label{eq:kdecdef}
k_{\rm dec}\simeq\left[{\rm erf}(y_d)-{\rm erf}(y_f)
+\frac{e^{-y_d^2}}{2\sqrt{\pi}y_d}\right]^{-1}\,,
\eeq
where $y\equiv \sqrt{x\epsilon_R}$ and $\rm{ erf}$ is the error function. The first two terms (last term) in the square brackets correspond to DM annihilation before (after) kinetic decoupling. In the regime $x_{d}\epsilon_R\ll 1$, the last term dominates, yielding a suppression of the relic density relative to the usual result without kinetic decoupling, 
\beq\label{eq:kdec}
k_{\rm dec}\simeq 2\sqrt{\pi x_d\epsilon_R}\ll 1\,.
\eeq
Once kinetic decoupling occurs, DM particles, being non-relativistic, cool more rapidly than the thermal bath, with $T'\propto T^2$. This  causes the resonance condition to be satisfied earlier, maximizing the annihilation cross-section when the Universe’s entropy density is still high and DM particles are less diluted. As a result, the DM yield decreases more rapidly, scaling approximately as $x^{-2}$. This effect weakens the later the kinetic decoupling occurs. 
In the opposite limit, $x_d\epsilon_R\gg 1$, the relic density remains unchanged, as expected, since the DM yield stops evolving after $x\sim \mathcal{O}(\epsilon_R^{-1})$. 
Interestingly, $k_{\rm dec}$ in~\cref{eq:kdecdef} reaches a maximum at finite $x_d=(2\epsilon_R)^{-1}$. This implies a window of enhanced relic density, though only by a modest factor of $\mathcal{O}(10\%)$ at most, with a mild $\epsilon_R$-dependence from the ${\rm erf} (y_f)$ term. The reason for this is that, at temperatures where the peak of the DM velocity distribution has moved below the resonance, annihilation becomes less efficient. This is because the distribution shifts away from the resonance more rapidly than in kinetic equilibrium, diminishing the overlap of the distribution's tail at later times. When decoupling occurs for $x_d\sim \epsilon_R^{-1}$, where the enhancement discussed earlier has largely diminished, this effect becomes dominant, leading to a slightly higher DM abundance. 
\Cref{fig:yield} shows the evolution of $Y_\chi$ with time for various kinetic decoupling temperatures.\\

We estimate the kinetic-decoupling temperature as follows. 
Kinetic equilibrium is maintained at least until chemical decoupling occurs at $x_f\approx 20$ thanks to the annihilation process itself~\cite{Belanger:2024bro}. Whether kinetic equilibrium persists to lower temperatures generally depends on the efficiency of scattering processes. While model-dependent, the generic properties of such processes can still be assessed. 
After chemical decoupling, the scattering rate $n_{\rm SM}\langle \sigma_{\rm scat.} v\rangle$ typically remains faster than Hubble expansion because the SM number density $n_{\rm SM}$ is not Boltzmann suppressed, ensuring kinetic equilibrium until well beyond freeze-out in models where the scattering and annihilation cross-sections are comparable. Here, however,  unlike annihilation, scattering is not resonantly enhanced and is expected to be suppressed by a factor of $\cO(\gamma_R)$ relative to annihilation. For sufficiently small resonance widths, the scattering rate is generically subdominant to the annihilation rate, so kinetic decoupling occurs immediately after chemical decoupling. In fact, as much as late resonant annihilation delays freeze-out, it also continues to heat DM particles after $x_f$, thereby extending the phase of kinetic decoupling to a significantly lower temperature $x_e\gg x_f$. As shown in \cref{app:kindec}, taking $x_d=x_f^{1-n}x_e^n$ with $n=4/7$ and  
\beq
x_e\simeq 2.7\times 10^{4}\left[\frac{b_R \gamma_R\bar{g}_\star^{1/2}\tilde{h}_{\rm eff}^{7/3}c}{\bar h_{\rm eff}m_{\chi\rm{GeV}}}\right]^{1/5}\epsilon_R^{-7/10}\,,
\eeq
where $c$ is a constant defined in \cref{eq:Cdef} and $\bar{h}_{\rm eff}$ and $\tilde{h}_{\rm eff}$ are the values of $h_{\rm eff}$ at $x_f$ and $x_e$, respectively, captures the effect of kinetic decoupling  within a factor of two. 
This conclusion may change in models where additional scattering processes, distinct from the crossing-symmetric counterparts of the annihilation processes, also contribute efficiently to maintain kinetic equilibrium beyond chemical decoupling. 
We disregard such processes in what follows.

\section{CMB constraints}\label{sec:CMBlimits} DM annihilation during recombination injects energy that heats and reionizes the photon-baryon plasma, affecting the CMB temperature and polarization fluctuations~\cite{Slatyer:2009yq}. Observations from the Planck satellite place stringent constraints on DM annihilation at that time~\cite{Planck:2018vyg}. In particular, for annihilation into electron pairs, CMB data  exclude cross-sections as large as $\langle \sigma v\rangle_{\rm th}= 3\times 10^{-26}\,{\rm cm}^3/{\rm s}$, for DM below $m_\chi\simeq 10\,$GeV~\cite{Slatyer:2015jla}. This typically rules out standard freeze-out via s-wave annihilation in this channel. However, we show that resonant annihilation can evade these bounds and determine the resonance parameters required for various final states, including electrons, muons, pions, and photons.

At recombination, DM particles are typically very cold, with $\epsilon_{\rm CMB}\ll 1$ when the Universe's temperature is $T_{\rm CMB}\approx0.26\,$eV. If $\epsilon_R\gg \epsilon_{\rm CMB}$, the annihilation cross-section at that time is non-resonant and well approximated by the zero-velocity limit ($\epsilon\to 0$) of ~\cref{eq:sigmav}, typically valid for $\epsilon_R\sim 10^{-5}$ (see~\cref{app:epsilonDM}). This gives, 
\begin{align}\label{eq:CMB}
\langle \sigma v\rangle_{\rm CMB}
=&\,\frac{8\pi b_R \gamma_R^2}{m_\chi^2\epsilon_R^{1/2}\left(\gamma_R^2+\epsilon_R^2\right)}\nonumber\\
\approx &\,
4.7\times 10^{-27}\,{\rm cm}^3/{\rm s}\,\frac{N_\chi x_d^{1/2}\gamma_R }{\bar g_\star^{1/2}f_\chi\epsilon_R^{3/2}}\,,
\end{align}
where we include in the second equality the effect of kinetic decoupling using~\cref{eq:kdec} in~\cref{eq:relicdensity} and assumed $\gamma_R\ll \epsilon_R$. The CMB bound can also be evaded if $\epsilon_R\ll \epsilon_{\rm CMB}$, but  since $\epsilon_{\rm CMB}\lesssim 10^{-11}$ for GeV-scale DM, this would require extreme fine-tuning of dark sector masses, so we do not consider this scenario further. 

An upper bound $\langle\sigma v\rangle_{\rm CMB}$ from CMB anisotropies constrains the resonance width, yielding
\beq
\gamma_R\lesssim 2.0\times 10^{-7}\,\frac{\bar{g}_\star^{1/2}f_\chi\langle\sigma v\rangle_{\rm CMB}}{N_\chi x_d^{1/2}\langle\sigma v\rangle_{\rm th}}\,\left[\frac{\epsilon_R}{10^{-5}}\right]^{3/2}\,,
\eeq
which suggests very weak resonance's couplings to DM and SM particles.\\

\begin{figure}
\includegraphics[width=0.5\textwidth]{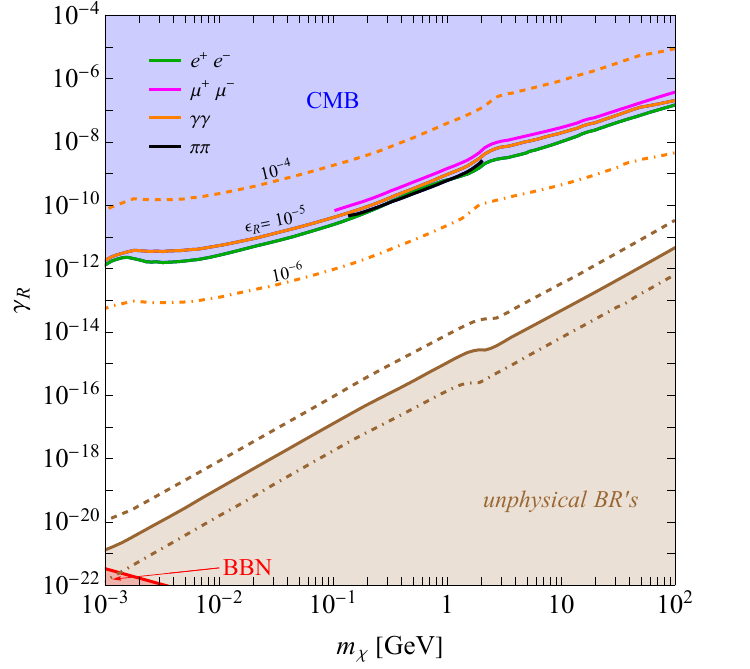}
\caption{Resonance width $\gamma_R$ as function of the DM mass $m_\chi$. The blue-shaded area is excluded by CMB observations with different boundaries representing annihilation to electrons (green), muons (magenta), photons (orange) and pions (black), assuming $\epsilon_R=10^{-5}$. The pion bound assumes annihilation into both $\pi^0\pi^0$ and $\pi^+\pi^-$ channels, assuming isospin symmetry. The red region corresponds to resonance lifetime exceeding one second, potentially affecting BBN predictions. In the brown-shaded region  the resonance branching ratios are not physical ($b_R>\omega/4$, assuming $J_R=1, S_\chi=0$). Dashed and dot-dashed lines represent results for resonances with $\epsilon_R=10^{-4}$ and $10^{-6}$, respectively, showing only photons for the CMB exclusion.}
\label{fig:gammaCMB}
\end{figure}

\Cref{fig:gammaCMB} shows the allowed resonance widths for DM annihilation into electrons, muons, pions, and photons as a function of the DM mass and for different values of $\epsilon_R$. 
We use {\tt micrOMEGAs} to derive the CMB constraint for the electron, muon, photon and pion channels, following the approach outlined in~\cite{Slatyer:2015jla} and accounting for the most recent Planck data~\cite{Planck:2018vyg}. The pion bound assumes annihilation into both $\pi^0\pi^0$ and $\pi^+\pi^-$ final states, with cross-sections related by isospin symmetry, though the neutral pion contribution largely dominates. Furthermore, for the interested reader, we provide CMB bounds for several individual channels in~\cref{app:CMBlimits}. 

The allowed region is bounded from below by the condition $\gamma_R\gtrsim 3.3\times 10^{-25}/m_{\chi{\rm GeV}}$ that the resonance lifetime does not exceed one second so that it has disappeared from the Universe before the onset of Big Bang nucleosynthesis (BBN), and by the condition $b_R<\omega/4$ corresponding to physically relevant resonance BRs. 
For electrons, assuming $\epsilon_R=10^{-5}$, the CMB constraint forces the resonance to be extremely narrow, as small as $\sim 10^{-12}$ of its mass for $m_\chi\approx 1\, $MeV. 

For DM below $10\,$MeV annihilating into electrons and photons, constraints from $N_{\rm eff}$ (not shown in~\cref{fig:gammaCMB}) may also be relevant (see, {\it e.g.},~\cite{Sabti:2019mhn,Sabti:2021reh,Chu:2022xuh}). However, these bounds are model-dependent, as they vary with the DM spin and can be significantly relaxed, in some cases down to $m_\chi\simeq1\,$MeV, if additional couplings to neutrinos are present~\cite{Sabti:2019mhn}.       

\section{Indirect detection signals}
\label{sec:IDsignals}

We now examine signals of DM annihilation in the present Universe, focusing on the galactic halo, dwarf galaxies, and galactic clusters. For resonant s-wave annihilation, the predicted cross-section depends  strongly on the DM velocity. We assume that DM particles have virialized within these astrophysical systems, following a Maxwell-Boltzmann distribution with a dispersion velocity $v_{\rm DM}$. 
Under the relic density constraint from~\cref{eq:relicdensity}, the velocity-averaged annihilation cross-section is given by,
\begin{align}
\label{eq:svhalo}
\langle \sigma v\rangle &\simeq 4.3 \times 10^{-22}\,{\rm cm}^3/{\rm s}\,\frac{N_\chi x_d^{1/2}}{\bar g_\star^{1/2}f_\chi}\left(\frac{v_\odot}{v_{\rm DM}}\right)^3\left[\frac{\epsilon_R}{10^{-5}}\right] \nonumber \\
&\quad \times \exp\left[-18.6\left(\frac{v_\odot}{v_{\rm DM}}\right)^2\left[\frac{\epsilon_R}{10^{-5}}\right]\right]\,,
\end{align}
where $v_\odot\equiv 220\,$km/s is the typical velocity dispersion in the Milky Way halo at a distance of $\sim10\,$kpc from the Galactic center~\cite{Freese:1987wu,Moffat:2024psv}.  Notably, $\langle\sigma v\rangle$ in~\cref{eq:svhalo} is independent of the resonance width $\gamma_R$ and only weakly depends on the DM mass via the number of relativistic degrees of freedom present in the early Universe at the time of DM chemical decoupling $\bar g_\star$. In astrophysical environments where $v_{\rm DM}$ is much smaller than $v_\odot$, such as dwarf galaxies, the cross-section is no longer resonantly enhanced and instead approaches the zero-velocity limit,
\begin{align}\label{eq:sv0}
\sigma v_0 \equiv \sigma v_{\epsilon \to 0} &\simeq 1.5 \times 10^{-31}\,{\rm cm}^3/{\rm s}\,\frac{N_\chi x_d^{1/2} }{\bar g_\star^{1/2} f_\chi} \nonumber\\
&\quad \times\left[\frac{\gamma_R}{10^{-12}}\right]\left[\frac{\epsilon_R}{10^{-5}}\right]^{-3/2}\,,
\end{align}
which is the same cross-section relevant during recombination. Additionally, $\sigma v_0$ depends weakly on $m_\chi$ through $\bar g_\star$.

\begin{figure}
\includegraphics[width=0.50\textwidth]{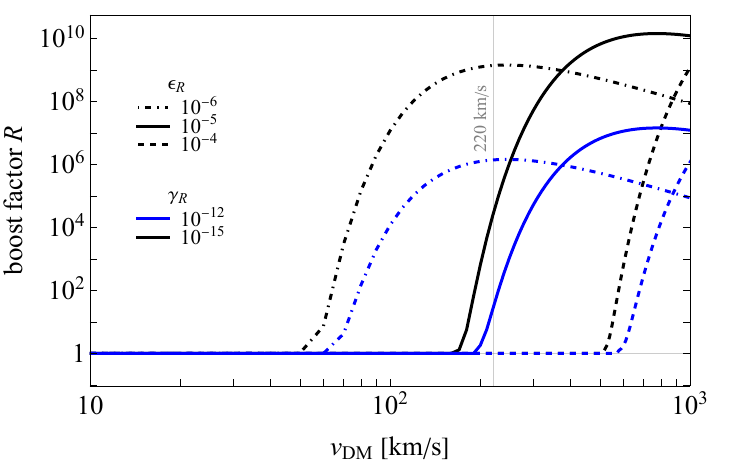}
\caption{Boost factor $R\equiv\langle \sigma v\rangle/\sigma v_0$  of the DM annihilation cross-section as function of the DM dispersion velocity $v_{\rm DM}$, relative to its zero-velocity counterpart, {\it i.e.}, the cross-section relevant during recombination. Dashed, solid and dot-dashed lines correspond to $\epsilon_R=10^{-4},\  10^{-5}$ and $10^{-6}$, respectively, with fixed $\gamma_R=10^{-12}$ (blue) and $10^{-15}$ (black). The vertical gray line marks the typical DM velocity dispersion $v_\odot=220\,$km/s in the Milky Way's halo, excluding the Galactic center~\cite{Freese:1987wu,Moffat:2024psv}.}
\label{fig:sv220}
\end{figure}

\Cref{fig:sv220} illustrates the boost factor $R\equiv \langle\sigma v\rangle/\sigma v_0$ due to BW effects as a function of the DM dispersion velocity, for different values of $\epsilon_R$ and $\gamma_R$, highlighting the linear dependence of $R$ on $\gamma_R$.\\

A variety of indirect detection (ID) experiments search for the products of DM annihilation in different astrophysical environments. These constraints typically assume a velocity-independent s-wave annihilation cross-section. However, near a resonance, the cross-section exhibits strong  velocity dependence, necessitating a recasting of experimental limits.

The Fermi-LAT telescope observes gamma-ray emissions from dwarf galaxies~\cite{Fermi-LAT:2015att}, where typical DM velocity dispersions range from  $2.5\,$km/s to $10.7\,$km/s~\cite{Zhao_2016}. At these low velocities, DM annihilation occurs far from the resonance, justifying the zero-velocity approximation $\langle \sigma v \rangle_{\text{Fermi-LAT}}\approx\sigma v_0$. Similarly, for constraints based on gas cooling observations of the Leo T dwarf galaxy~\cite{Wadekar:2021qae}, where the DM  velocity is estimated  to be around $7\,$km/s~\cite{Ryan-Weber:2007guk,Adams_2018}.

Additional constraints can be extracted from  X-ray data of XMM-Newton~\cite{Cirelli:2023tnx} and eROSITA~\cite{Balaji:2025afr}, which observe the entire galaxy and derive limits from concentric rings. We use the constraint from the $42^\circ-48^\circ$ latitude region, which provides to the strongest limits away from the galactic center (GC). In this region, the dominant contribution to DM annihilation comes from DM particles with a velocity $v_\odot$.
The X-ray constraints are subject to large uncertainties; here, we adopt the central value from~\cite{Cirelli:2023tnx}.

\begin{figure}[t]
\centering
\begin{tabular}{c}
\includegraphics[width=0.5\textwidth]{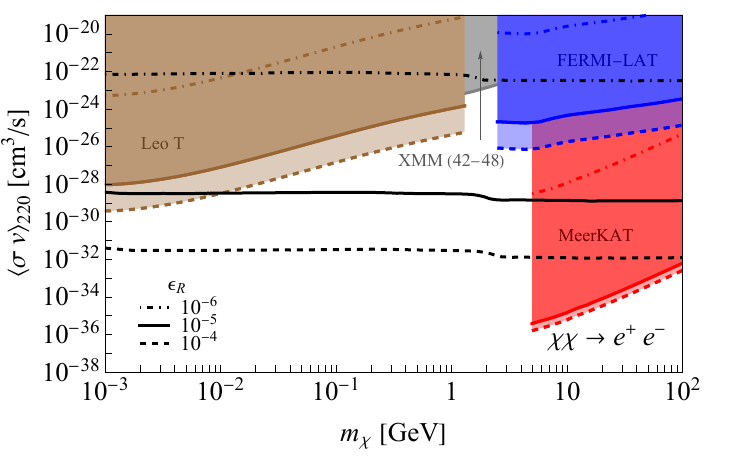}\\
\includegraphics[width=0.5\textwidth]{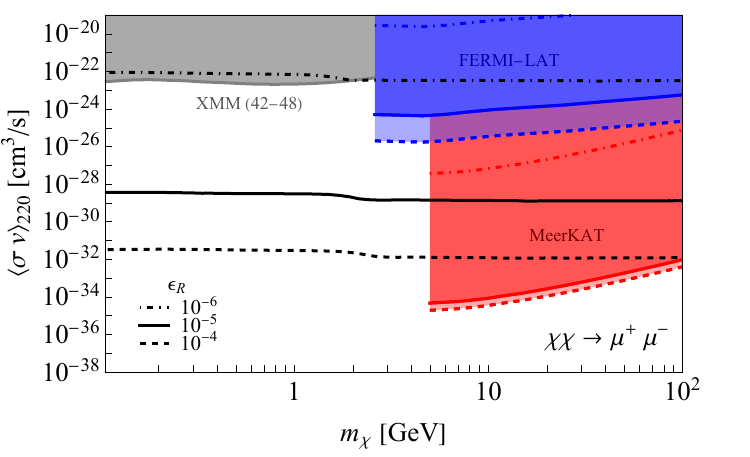}
\end{tabular}
\caption{DM annihilation cross-section $\langle\sigma v\rangle_{220}$ as function of the DM mass $m_\chi$, predicted by the relic density for a typical velocity dispersion $v_\odot=220\,$km/s and $\gamma_R=10^{-12}$ (black lines). The shaded regions indicate exclusions (recast as described in~\cref{sec:IDsignals}) from previous DM indirect detection searches in the $e^+e^-$ (top) and $\mu^+\mu^-$ (bottom) channels: Fermi-LAT observations of Milky Way dwarf spheroidal galaxies~\cite{Fermi-LAT:2015att} (blue), constraints from the gas-rich dwarf galaxy Leo T~\cite{Wadekar:2021qae} (brown), the galaxy cluster legacy survey by MeerKAT~\cite{Lavis:2023jju} 
and X-ray observations a latitude ring between $42^\circ$ and $48^\circ$ by XMM-Newton~\cite{Cirelli:2023tnx} (gray). Dashed, solid and dot-dashed lines correspond to $\epsilon_R=10^{-4},\,10^{-5}$ and $10^{-6}$, respectively.}
\label{fig:ID}
\end{figure}

We also incorporated radio constraints from the MeerKAT telescope, which detects synchroton emissions from galaxy clusters and dwarf galaxies using radio interferometry~\cite{Knowles:2021nvz,Beck:2023oza}. We focus on constraints from galaxy clusters, which have fewer astrophysical uncertainties and provide more robust bounds~\cite{Lavis:2023jju}. Using {\tt DarkMatters}~\cite{Sarkis:2024zjg}, we computed the radio flux from DM annihilation into electrons and muons and compared them to MeerKAT's L-band sensitivity~\cite{Jonas:2018Jr}. The typical DM velocity in these clusters is $\sim 1000\,$km/s, resulting in a large boost factor $R$ (see~\cref{fig:sv220}) for a wide range of $\epsilon_R$. The constraints in~\cref{fig:ID} are derived from observations of the  Abell 133 cluster~\cite{Lavis:2023jju}. These limits depend on astrophysical parameters, such as the halo radius, density profile, and magnetic field, and can weaken by a factor of 2, or strengthen by over an order of magnitude. 

Other constraints from both INTEGRAL~\cite{Cirelli:2020bpc} and  COMPTEL~\cite{Essig:2013goa} are derived from the inner region of the galaxy, but due to uncertainty in DM velocity in this region, a detailed recasting is required, which we do not perform. \\

\Cref{fig:ID} presents the predicted annihilation cross-section $\langle \sigma v\rangle_{220}\equiv \langle \sigma v\rangle(v_{\rm DM}=v_\odot)$  that satisfies the relic density constraint for different values of $\epsilon_R$, compared to the limits extracted from XMM-Newton ($m_\chi<5\,{\rm GeV}$), Leo T ($m_\chi<1.3\,{\rm GeV}$), Fermi-LAT (for $m_\chi>2\,{\rm GeV}$), and MeerKAT ($m_\chi>5\,{\rm GeV}$), assuming annihilation into $e^+e^-$ (top panel) or
$\mu^+\mu^-$ (bottom panel), with $\gamma_R=10^{-12}$. For $\epsilon_R=10^{-6}$, the maximum resonant enhancement occurs at $v_\odot$, leading to strong XMM-Newton and MeerKAT constraints that exclude DM over the entire mass range. 
For $\epsilon_R=10^{-5}$, XMM-Newton provides the most stringent constraint for $1\ {\rm GeV}\lesssim m_\chi<5\ {\rm GeV}$, but does not exclude the relic-density compatible region. Notably, due to resonance effects, very small couplings are required to reproduce the observed DM abundance, leading to $\langle \sigma v\rangle_{220}$ values  much smaller than $3\times 10^{-26} {\rm cm}^3/s$. MeerKAT rules out the region $m_\chi > 5\ {\rm GeV}$, as it benefits from a large boost factor. The same conclusion applies to both leptonic channels.
For $\epsilon_R=10^{-4}$, the resonance peak shifts to higher velocities, suppressing $\langle \sigma v \rangle_{220}$ and allowing sub-GeV DM to evade current constraints, while MeerKAT excludes DM in the 5--100$\,$GeV range in both leptonic channels. 
Due to the small DM velocities involved, dwarf galaxies observations do not benefit from a resonance boost and thus fail to constrain  resonant DM for any  $\epsilon_R$. Similar conclusions hold for DM annihilation into pion pairs. 

\section{Conclusion}\label{sec:conclusions}
In this work, we analyzed the phenomenology of s-wave resonant DM annihilation using a model-independent BW approach. We  focused on resonance parameters that significantly enhance DM annihilation during freeze-out, allowing for the correct relic density while evading stringent constraints from CMB data. This mechanism  revives the thermal s-wave DM scenario for masses below $\sim 10\,$GeV.

The BW effects in DM annihilation enables thermal DM production with much weaker DM-to-SM interactions compared to the standard freeze-out scenario. This leads to early kinetic decoupling during freeze-out and a strong suppression of the relic density. As a result, during recombination, when DM particles are extremely cold, annihilation occurs off-resonance and can be well below the thermal expectation. However, evading current CMB constraints requires the existence of extremely narrow states in the dark sector, with  widths in the range $10^{-16}\lesssim\gamma_R\lesssim 10^{-9}$ for $m_\chi=1\,$GeV. This suggests the presence of high-quality approximate dark symmetries.  

We also investigated constraints and prospects from indirect detection experiments, which probe DM annihilation across different velocity regimes in the galactic halo, dwarf spheroidal galaxies, and galaxy clusters using data from XMM-Newton, Leo T, Fermi-LAT, and MeerKAT. Among these, MeerKAT is particularly effective in further constraining the resonance parameter space, while still  leaving room for potential DM annihilation signals in future searches below the electroweak scale. However, future advancements in indirect detection experiments would only improve the sensitivity to the $\epsilon_R$ parameter logarithmically.

\section*{Acknowledgments}

We thank Alexander Pukhov for his help in improving the spectra tables in {\tt micrOMEGAs} and Geoff Beck for helping with {\tt DarkMatters}. We also thank Marco Cirelli and Jordan Koechler for providing X-ray exclusions for intermediate latitudes. Finally, M.~J. thanks Pierre Salati for useful, enthusiastic discussions. S.~C. also acknowledges support from the UKRI Future Leader Fellowship “DARKMAP”
(Grant No: MR/T042575/1). 

\appendix
\section{Freeze-out near a resonance}\label{app:resonantFO}
We briefly review here the DM relic density's determination for annihilation near a resonance, following Ref.~\cite{Gondolo:1990dk}. The thermal average of the annihilation cross-section times velocity in~\cref{eq:sigmav},
in the non-relativistic limit and at leading order in $\epsilon_R\ll 1$, yields,
\beq\label{eq:sigmavavg}
\langle \sigma v\rangle=\frac{16\pi^{3/2}b_R\gamma_R}{m_\chi^2}x^{3/2}e^{-x\epsilon_R}\,,
\eeq
where $x\equiv m_\chi/T$ and we used the narrow-width approximation, $\gamma_R/[(\epsilon-\epsilon_R)^2+\gamma_R^2]\to \pi \delta(\epsilon-\epsilon_R)$ when $\gamma_R\to 0$. 
After chemical decoupling at $x=x_f$, $n_\chi$ no longer tracks $n_{\chi\,{\rm eq}}$ which becomes negligible in~\cref{eq:BEq}. Hence,
the DM yield $Y_\chi\equiv n_\chi/s$, with $s=2\pi^2 h_{\rm eff}(T)T^3/45$ the Universe's entropy density, solving~\cref{eq:BEq} is approximately~\cite{Gondolo:1990dk},
    \beq\label{eq:Ysol}
Y_\chi^{-1}(x)= Y_f^{-1}+
\left(\frac{45}{\pi}G\right)^{-1/2}m_\chi\int_{x_f}^x dx' \, \frac{g_\star^{1/2}\langle \sigma v\rangle}{x^{\prime\,2}}\,,
\eeq
where $G\equiv M_{\rm Pl}^{-2}$ is Newton's constant expressed in units of the Planck's mass $M_{\rm Pl}\approx 1.2\times 10^{19}\,$GeV, 
\beq
g_\star^{1/2}\equiv \frac{h_{\rm eff}}{g_{\rm eff}^{1/2}}\left(1+\frac{d\log h_{\rm eff}}{3\,d\log T}\right)\,,
\eeq
and $Y_f\equiv Y_{\chi\,\rm eq}(x_f)$. Using~\cref{eq:sigmavavg}  yields, for $x\gg x_f$,
\beq\label{eq:Yofx}
Y_\chi(x)\simeq  \frac{3 m_\chi\sqrt{5} \epsilon_R^{1/2}}{16\pi^{5/2}M_{\rm Pl} b_R\gamma_R \bar g_\star^{1/2}{\rm erf}\left(\sqrt{x\epsilon_R}\right)}\,,
\eeq
where we neglected the contribution from $x_f$ and the temperature dependence of $g_\star^{1/2}$, fixing it instead to $\bar g_\star^{1/2}\equiv g_\star^{1/2}(x_f)$. The relic abundance today is,
\beq\label{eq:Omega}
\Omega_\chi h^2 = 2.8\times 10^8\frac{N_\chi m_\chi}{\rm GeV}Y_\chi(x\to \infty) \,,
\eeq
where $N_\chi= 1$ ($N_\chi=2$) when DM particles and antiparticles are (not) identical. Combining~\cref{eq:Omega,eq:Yofx} yields~\cref{eq:Omchi} in the main text.    

Note that the DM yield evolves long after chemical decoupling, until $x\sim\cO(\epsilon_R^{-1})\gg x_f$, which is in contrast with canonical thermal DM whose yield freezes out soon after $x_f$. The reason is that $\langle\sigma v\rangle$ in~\cref{eq:sigmavavg} increases faster at lower temperatures due to the resonance. After chemical decoupling ($x>x_f$), DM particles get colder due to expansion, thus enhancing the probability for two particles to satisfy the resonance condition. In this regime, the DM yield approximately scales as $x^{-1/2}$. Once $x\gtrsim \epsilon_R^{-1}$, particles become too slow to annihilate resonantly and DM freezes out. However, this simple picture is significantly altered by kinetic decoupling which typically occurs beforehand.

\section{Resonant freeze-out with kinetic decoupling}\label{app:kdec}
In the presence of kinetic decoupling, assuming $n_\chi\gg n_{\chi\,{\rm eq}}/\sqrt{\beta}$ after chemical decoupling, the DM yield is still given by~\cref{eq:Ysol} but with the thermal average replaced by $\langle \sigma v\rangle'=\beta\langle\sigma v\rangle$ with $\beta$ given by~\cref{eq:beta}. Repeating the steps above, we find, 
\beq\label{eq:Ycorrected}
Y_\chi(x)=Y_\chi^{\rm keq}(x)F(\sqrt{x\epsilon_R})\,,
\eeq
where $Y_\chi^{\rm keq}$ denotes the solution in~\cref{eq:Ysol} obtained assuming kinetic equilibrium. The function $F(y<y_d)=1$ and,
\beq\label{eq:Fofy}
F(y>y_d)\simeq\frac{{\rm erf}(y)-{\rm erf}(y_f)}{{\rm erf}(y_d)-{\rm erf}(y_f)+\frac{e^{-y_d^2}-e^{-y^4/y_d^2}}{(2\sqrt{\pi}y_d)}}\,.
\eeq
The $\chi$'s abundance drops faster after kinetic decoupling, scaling approximately as $x^{-2}$.

\section{DM average velocity at recombination}
\label{app:epsilonDM}

The average velocity of DM after chemical decoupling is,
\beq
\epsilon(T\leq T_f)= \frac{3T'(T)}{2m_\chi}\,,
\eeq
where the DM temperature $T'$ equals the plasma temperature $T$ until kinetic decoupling at $T_d$ after which it evolves as in~\cref{eq:Tprime}. Kinetic decoupling typically occurs well before recombination, $T_{\rm CMB}\approx 0.26\,{\rm eV}\ll T_d$. Therefore,
\beq
\epsilon_{\rm CMB}\equiv \epsilon(T_{\rm CMB})=\frac{3T_{\rm CMB}^2 r_{df}}{2m_\chi^2 x_f}\,.
\eeq
Using $x_f=20$, we obtain $\epsilon_{\rm CMB}\simeq 2.6\times 10^{-13}\gamma_R/ m_{\chi{\rm GeV}}^2$ for $\gamma_R>\gamma_R^c$, and $\epsilon_{\rm CMB}\simeq 4.8\times 10^{-18}/ m_{\chi{\rm GeV}}^2$ otherwise. DM is less cold in the former case due to later kinetic decoupling. Restricting to narrow resonances ($\gamma_R<\epsilon_R\ll 1$) then yields an upper limit on DM velocity at recombination,
\beq
\epsilon_{\rm CMB}\lesssim 2.6\times 10^{-13}m_{\chi{\rm GeV}}^{-2}\,,
\eeq
which is smaller than $10^{-7}$ unless $m_\chi\lesssim 2\,$MeV.

\section{Kinetic decoupling analysis}
\label{app:kindec}

We follow here the approach of~\cite{Binder:2017rgn,Binder:2021bmg} and solve the system of coupled Boltzmann equations (cBE) governing the evolution of $Y_\chi$ and the DM temperature represented by the dimensionless quantity $y_\chi\equiv m_\chi T'/s^{2/3}$,
\begin{align}
 \frac{d\log Y_\chi}{dx}&=\frac{sY_\chi}{x\tilde H}\left[\frac{Y_{\chi\,\rm{eq}}^2}{Y_\chi^2}\langle\sigma v\rangle-\langle\sigma v\rangle' \right]\,,\label{eq:cBEY}\\
\frac{d\log y_\chi}{dx}&= \frac{\gamma_{\rm sca}}{x \tilde H}\left[\frac{y_{\chi\,\rm{eq}}}{y_\chi}-1\right]+\frac{s Y_\chi}{x\tilde H}\Big[\langle \sigma v\rangle'-\langle\sigma v\rangle'_2\Big]\nonumber\\
&\quad+\frac{sY_\chi}{x \tilde H}\frac{Y_{\chi\,\rm{eq}}^2}{Y_\chi^2}\left[\frac{y_{\chi\,\rm{eq}}}{y_\chi}\langle\sigma v\rangle_2-\langle\sigma v\rangle\right] \nonumber\\
&\quad+\frac{y_{\chi\,\rm{eq}}}{ y_\chi}\frac{H}{\tilde H}\left\langle \frac{p^4}{3m_\chi E^3}\right\rangle'\,,\label{eq:cBEy}
\end{align}
where primed averages are evaluated at temperature $T'$. $\langle\sigma v\rangle_2$ is the temperature-weighted thermal average of the annihilation cross-section~\cite{Binder:2017rgn}, which for resonant annihilation is related to $\langle \sigma v\rangle$ as
\begin{align}
\frac{\langle\sigma v\rangle_2}{\langle\sigma v\rangle}&=\frac{2x^2}{3K_1(2x)}\int_1^\infty d\epsilon_+\frac{(\epsilon_+^2-1)^{3/2}}{\epsilon_+}e^{-2x\epsilon_+}
\end{align}
In the non-relativistic limit, the integrand sharply peaks close to $\epsilon_+=1$, yielding
\beq
\langle\sigma v\rangle_2\simeq\frac{\langle\sigma v\rangle}{2}\left(1-\frac{1}{2x}+\cdots\right)\,,
\eeq
where $\cdots$ denote higher powers of $x^{-1}\ll 1$. Moreover, $\langle p^4/(3m_\chi E^3)\rangle\simeq 5/x^2$ is negligible in this limit. The further neglect of the scattering rate $\gamma_{\rm sca}$, which is $\mathcal{O}(\gamma_R^{-1})$-suppressed relative to the annihilation rate, and terms $\propto Y_{\chi\,\rm{eq}}^2/Y_\chi^2$, which are exponentially suppressed after chemical decoupling, the cBE system in \cref{eq:cBEY,eq:cBEy} simplifies to
\beq
\frac{d\log Y_\chi}{dx}\simeq -2\frac{d\log y_\chi}{dx}\simeq -\frac{s Y_\chi}{x \tilde H}\langle\sigma v\rangle'\,,
\eeq
showing that the DM yield and temperature have nearly parallel evolution with $Y_\chi(x)y_\chi^2(x)$ being approximately constant for $x\gg 1$.\\
\begin{figure}[t]
    \centering
   \begin{tabular}{c}
    \includegraphics[width=0.5\textwidth]{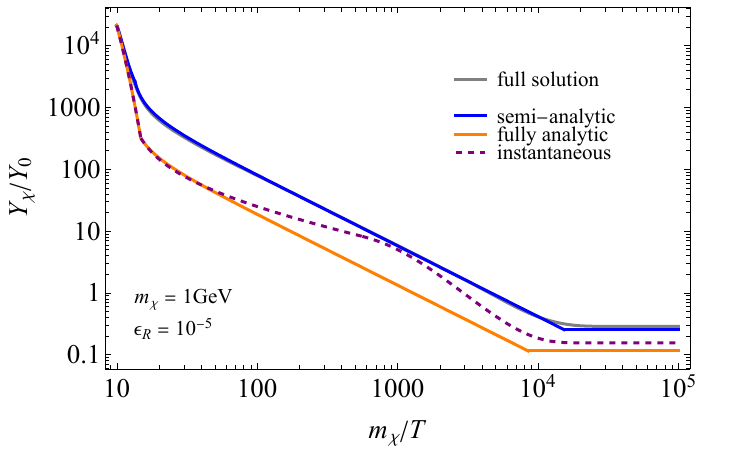} \\
    \includegraphics[width=0.5\textwidth]{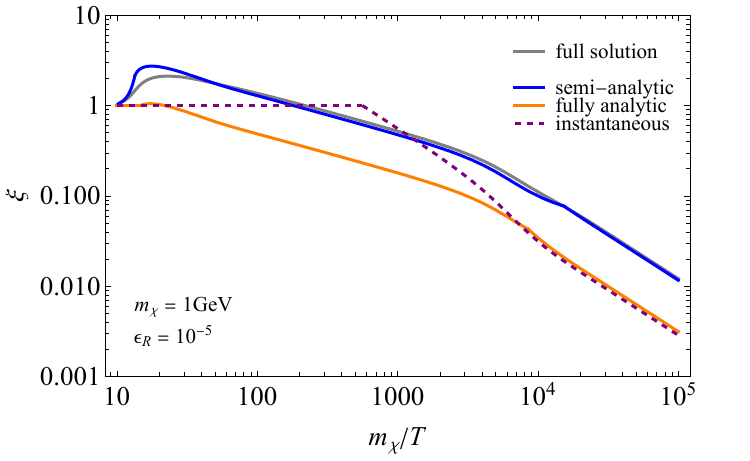}
   \end{tabular}
    \caption{DM abundance $Y_\chi$ (top) and temperature ratio $\xi=T'/T=y_\chi/y_{\chi\,\rm{eq}}$ (bottom) as functions of $x=m_\chi/T$, from a full numerical solution of the cBE system (gray), the semi-analytic approximation (blue), the fully analytic approximation (orange), and the simplified analytic prescription with instantaneous kinetic decoupling (dashed purple). Abundances are shown in units of the kinetic equilibrium value at $x\to\infty$, denoted as $Y_0$.}
    \label{fig:comparison}
\end{figure}
\paragraph{Approximate solution at large $x$} At sufficiently large $x$, the DM yield satisfies $Y_\chi(x)\simeq c/y_\chi^2(x)$, where $c$ is a constant. In this regime,  the cBE system is effectively decoupled, and $y_\chi$ obeys
\beq\label{eq:yXeq}
y_\chi^{5/2}\frac{dy_\chi}{dx}=\frac{c \sigma_0 m_\chi^3}{2x \tilde H}\exp\left[-\frac{\epsilon_R m_\chi^2}{y_\chi s^{2/3}}\right]\,,
\eeq
with $\sigma_0\equiv 16\pi^{3/2}b_R \gamma_R/m_\chi^2$. For $y_\chi\ll \epsilon_R m_\chi^2/s^{2/3}$, the exponential tends to unity. Ignoring the variation in the number of degrees of freedom, one finds      
\beq
y_\chi(x)\simeq \left[a(x^2-x_f^2)-\bar{y}_\chi^{7/2}\right]^{2/7}\,,
\eeq
where 
\beq
a\equiv \frac{21\sqrt{5}}{16\pi^{3/2}}c\sigma_0m_\chi M_{\rm Pl}\frac{\bar{g}_\star^{1/2}}{\bar{h}_{\rm eff}}\,,
\eeq
and barred quantities are evaluated at $x_f$, which we identify as the temperature at which kinetic decoupling begins. For $x\gtrsim x_f$, $y_\chi$ approximately grows like $x^{4/7}$ until the temperature 
\beq\label{eq:xe}
x_e\simeq \left(\frac{4\sqrt{2}\pi^2a^{3/7}\tilde h_{\rm eff}}{45\epsilon_R^{3/2}}\right)^{7/15}\,,
\eeq
where $\tilde h_{\rm eff}$ the value of $h_{\rm eff}$ at $x_e$. At this point, the exponent in \cref{eq:yXeq} become of order unity. For $x\gtrsim x_e$, the annihilation rate is exponentially suppressed, and $y_\chi$ approaches a constant, marking the end of the kinetic-decoupling phase. In this region, the DM yield behaves as $c/y_\chi^2\propto x^{-8/7}$, and its present day value is approximately
\beq
Y_\chi(x\to \infty)\simeq c \left(\frac{45\epsilon_R^{3/2}}{4\sqrt{2}\pi^2a^{3/2}\tilde h_{\rm eff}}\right)^{8/15}\,,
\eeq
which scales like $\sigma_0^{-4/5}$.\\

\paragraph{Near equilibrium solutions}
A reliable numerical determination of the chemical-decoupling time $x_f$ and the constant $c$ can be obtained by expanding the cBE system in small perturbations about equilibrium. Define $U_\chi(x)\equiv Y_\chi(x)/Y_{\chi\,\rm{eq}}(x)-1$ and $u_\chi(x)\equiv y_\chi(x)/y_{\chi\,\rm{eq}}(x)-1$. Near $x=x_f\sim\mathcal{O}(20)$ and to leading order in $U_\chi$ and $u_\chi$, \cref{eq:cBEY,eq:cBEy} reduce to (neglecting the scattering term)   
\begin{align}
\frac{dU_\chi}{dx}&-\left(1-\frac{3}{2x}\right)(1+U_\chi)= \frac{sY_{\chi\,\rm{eq}}}{x\tilde H}\langle\sigma v\rangle\left[\frac{3u_\chi}{2}-2U_\chi\right]\,,\\
\frac{du_\chi}{dx}&+\frac{1}{x}(1+u_\chi) =\frac{sY_{\chi\,\rm{eq}}}{x\tilde H}\langle\sigma v\rangle\left[\left(1+\frac{1}{2x}\right)U_\chi\right.\nonumber\\
&\left.\quad-\frac{5u_\chi}{4}\left(1-\frac{1}{10x}\right)+u_\chi U_\chi\left(1-\frac{1}{2x}\right)\right]\,.
\end{align}
Solving this system yields a numerical estimate of $x_f$, which is defined conventionally by $U_\chi(x_f)=\delta$ where $\delta=3/2$~\cite{Gondolo:1990dk}. This value of $x_f$ then allows to compute a numerical estimate of the constant $c=Y_\chi(x_f)y_\chi^2(x_f)$. \\

\paragraph{Analytical estimate of $x_f$}
For $x\gg 1$ and prior to chemical decoupling, $U_\chi=u_\chi=0$. In this limit we have approximately $dU_\chi/dx\simeq 1$ and $du_\chi/dx\simeq 0$, indicating that $U_\chi$  grows more rapidly than $u_\chi$ and that chemical decoupling occurs when $u_\chi$ is still relatively small. Consequently, a reliable analytical estimate of $x_f$ can be obtained under the assumption of kinetic equilibrium. Following the standard derivation in~\cite{Gondolo:1990dk}, one finds
\begin{align}\label{eq:xfanalytic}
x_f&\simeq \log\left(\frac{3\sqrt{5/2}g_\star^{1/2}}{4\pi^3 h_{\rm eff}}g_\chi m_\chi M_{\rm Pl}\sigma_0\delta(\delta+2)\right)\nonumber\\
&\quad+\log\left[\log\left(\frac{3\sqrt{5/2}g_\star^{1/2}}{4\pi^3 h_{\rm eff}}g_\chi m_\chi M_{\rm Pl}\sigma_0\delta(\delta+2)\right)\right]\,,
\end{align}
where $g_\chi$ denotes the number of helicity states of $\chi$. This analytic expression reproduces the numerically computed value of $x_f$ to within a few percent.\\

\paragraph{Comparison of approximate solutions}
We consider several approximations for determining the relic density while accounting for the impact of kinetic decoupling: {\it i) Semi-analytic approach}: $Y_\chi$ and $y_\chi$ are solved numerically up to $x_f$ using the linearized cBE system and then matched to the analytical approximate solutions at large $x$; {\it ii) Fully analytic approach}: Both $Y_\chi$ and $y_\chi$ are assumed to follow their equilibrium values up to $x_f$, given by \cref{eq:xfanalytic}, with
\beq\label{eq:Cdef}
c\simeq Y_{\chi\,\rm{eq}}(x_f)y^2_{\chi\,\rm{eq}}(x_f)\,;
\eeq
 {\it iii) Simplified analytic prescription}: An instantaneous kinetic decoupling is assumed at a temperature $x_f<x_d<x_e$, chosen  so that the $x^{-1/2}$ scaling of $Y_\chi$ for $x<x_d$ (see \cref{eq:Yofx}), followed by a steeper $x^{-2}$ decrease (see \cref{eq:Ycorrected,eq:Fofy}),  matches the approximate $x^{-8/7}$ scaling of the full cBE solution. This gives $x_d=x_f^{1-n}x_e^n$, with $n=4/7$, and $x_f$ and $x_e$ given by \cref{eq:xfanalytic,eq:xe}, respectively. This is the approach used to determine the relic density in the main text.\\

\Cref{fig:comparison} compares the DM abundance and temperature obtained with these three approaches for $\epsilon_R=10^{-5}$ and $m_\chi=1\,$GeV to the full numerical cBE solutions. In particular, the simplified analytical prescription underestimates the abundance of the full numerical solution by $\mathcal{O}(40\%)$. Moreover, for a representative set of benchmark points spanning the entire parameter space of interest, we verified that the difference between this approach and  cBE does not exceed a factor of two. 

\begin{figure}[t]
    \centering
   \begin{tabular}{c}
    \includegraphics[width=0.5\textwidth]{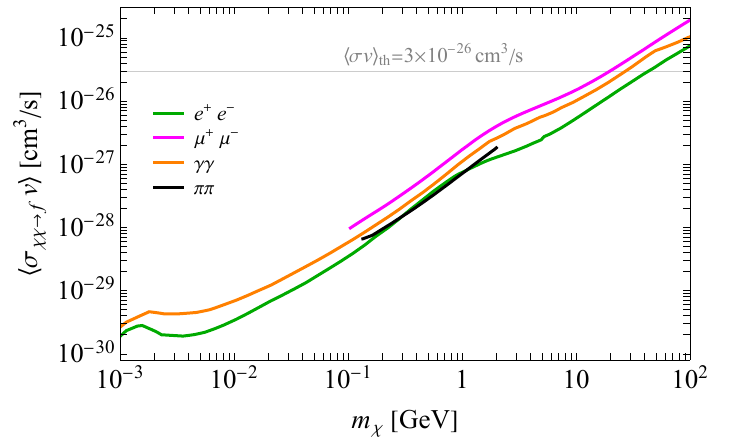} \\
    \includegraphics[width=0.5\textwidth]{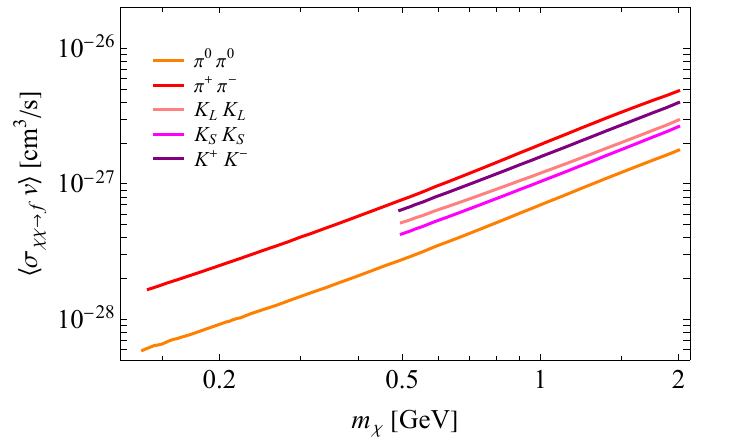}
   \end{tabular}
    \caption{CMB limits from Planck data for DM annihilation into $e^+e^-$, $\mu^+\mu^-$, $\gamma\gamma$ and $\pi\pi$ channels (top) and individual hadronic channels (bottom). For the $\pi\pi$ channel we present the limit on the total annihilation cross-section, which includes both the $\pi^0\pi^0$ and $\pi^+\pi^-$ final states, assuming isospin symmetry.}
    \label{fig:CMB-bounds}
\end{figure}
%

\section{CMB limit computation}
\label{app:CMBlimits}
Our CMB limits are derived with {\tt micrOMEGAs}~\cite{Alguero:2023zol} following the approach of~\cite{Slatyer:2009yq,Slatyer:2015jla} and using the latest Planck constraint~\cite{Planck:2018vyg} on the function 
\beq
P_{\rm ann}<3.2\times 10^{-28}\,{\rm cm}^3/{\rm s}/{\rm GeV}\,,
\eeq
at 95$\%$ confidence level (CL). $P_{\rm ann}$ is related to the annihilation cross-section through the following formula, $P_\text{ann}=f_{\text{eff}}\langle \sigma v \rangle/m_\chi$, where
\begin{equation}
    f_{\text{eff}}=\frac{\int_0^{m_\chi}EdE\left[ 2f_{\text{eff}}^{e+e-}(E)\left(\frac{dN}{dE}\right)_{e^+}+ f_{\text{eff}}^{\gamma}(E)\left(\frac{dN}{dE}\right)_{\gamma}  \right]}{2 m_\chi},
\end{equation}
is the efficiency factor with which the DM rest-mass energy released by annihilation is injected in the intergalactic medium and $(dN/dE)_{e^+,\gamma}$ are the energy spectra of positrons and photons, respectively, produced by the annihilation products. We used  {\tt micrOMEGAs} to compute  the spectra for electron, muon, photon and scalar mesons, down to sufficiently low energies to derive limits for DM masses as small as $1\,$MeV.  We have verified that the positron and photon spectra are in good agreement with {\tt Hazma}~\cite{Coogan_2020,Coogan:2022cdd} as well as~\cite{Gonzalez-Morales:2017jkx} for electron, muon and $\pi^\pm,\pi^0$ annihilation products, for $m_\chi\lesssim1\,$GeV.  In addition, we computed the spectra for the $K_S, K_L, K^\pm,\eta$ and $\eta'$ scalar mesons and included them in {\tt micrOMEGAs}.

The resulting CMB limits for individual annihilation channels are shown in~\cref{fig:CMB-bounds}. 
For $m_\chi\gtrsim 5\,$GeV, our limits for the $e^+e^-$, $\mu^+\mu^-$ and $\gamma\gamma$ channels agree within a few percent with those of~\cite{Slatyer:2015jla} derived using {\tt PPPC4DMID}~\cite{Cirelli:2010xx}, after rescaling the latter by a factor of $3.2/4.1$ to account for the latest Planck constraint. Our limits for the $\pi^0\pi^0$ 
channel is also in good agreement (below 5\%) with~\cite{Gonzalez-Morales:2017jkx}, while for the $\pi^+\pi^-$ channel our limit  is stronger by a factor of  2 near threshold and in good agreement at 1 GeV.

\bibliography{DMres}

@article{Balaji:2025afr,
    author = "Balaji, Shyam and Cleaver, Damon and De la Torre Luque, Pedro and Michailidis, Miltiadis",
    title = "{Dark Matter in X-rays: Revised XMM-Newton Limits and New Constraints from eROSITA}",
    eprint = "2506.02310",
    archivePrefix = "arXiv",
    primaryClass = "hep-ph",
    reportNumber = "KCL-PH-TH/2025-19",
    month = "6",
    year = "2025"
}

@article{Cheng:2023dau,
    author = "Cheng, Yu and Ge, Shao-Feng and Sheng, Jie and Yanagida, Tsutomu T.",
    title = "{Dark matter annihilation via Breit-Wigner enhancement with heavier mediator}",
    eprint = "2309.12043",
    archivePrefix = "arXiv",
    primaryClass = "hep-ph",
    doi = "10.1016/j.physletb.2025.139290",
    journal = "Phys. Lett. B",
    volume = "861",
    pages = "139290",
    year = "2025"
}

@article{Berlin:2023qco,
    author = "Berlin, Asher and Krnjaic, Gordan and Pinetti, Elena",
    title = "{Reviving MeV-GeV indirect detection with inelastic dark matter}",
    eprint = "2311.00032",
    archivePrefix = "arXiv",
    primaryClass = "hep-ph",
    reportNumber = "FERMILAB-PUB-21-457-T",
    doi = "10.1103/PhysRevD.110.035015",
    journal = "Phys. Rev. D",
    volume = "110",
    number = "3",
    pages = "035015",
    year = "2024"
}

@article{Moffat:2024psv,
    author = "Moffat, J. W. and Sharron, H. and Toth, V. T.",
    title = "{Implications of the Milky Way Declining Rotation Curve}",
    eprint = "2409.17371",
    archivePrefix = "arXiv",
    primaryClass = "astro-ph.GA",
    month = "9",
    year = "2024"
}

@article{Cirelli:2010xx,
    author = "Cirelli, Marco and Corcella, Gennaro and Hektor, Andi and Hutsi, Gert and Kadastik, Mario and Panci, Paolo and Raidal, Martti and Sala, Filippo and Strumia, Alessandro",
    title = "{PPPC 4 DM ID: A Poor Particle Physicist Cookbook for Dark Matter Indirect Detection}",
    eprint = "1012.4515",
    archivePrefix = "arXiv",
    primaryClass = "hep-ph",
    reportNumber = "CERN-PH-TH-2010-057, SACLAY-T10-025, IFUP-TH-2010-44",
    doi = "10.1088/1475-7516/2012/10/E01",
    journal = "JCAP",
    volume = "03",
    pages = "051",
    year = "2011",
    note = "[Erratum: JCAP 10, E01 (2012)]"
}

@article{Wang:2025tdx,
    author = "Wang, Yu-Ning and Duan, Xin-Chen and Tang, Tian-Peng and Wang, Ziwei and Tsai, Yue-Lin Sming",
    title = "{Exploring sub-GeV dark matter via $s$-wave, $p$-wave, and resonance annihilation with CMB data}",
    eprint = "2502.18263",
    archivePrefix = "arXiv",
    primaryClass = "hep-ph",
    month = "2",
    year = "2025"
}

@article{Ibe:2008ye,
    author = "Ibe, Masahiro and Murayama, Hitoshi and Yanagida, T. T.",
    title = "{Breit-Wigner Enhancement of Dark Matter Annihilation}",
    eprint = "0812.0072",
    archivePrefix = "arXiv",
    primaryClass = "hep-ph",
    reportNumber = "SLAC-PUB-13479, IPMU08-0097",
    doi = "10.1103/PhysRevD.79.095009",
    journal = "Phys. Rev. D",
    volume = "79",
    pages = "095009",
    year = "2009"
}

@article{Srednicki:1988ce,
    author = "Srednicki, Mark and Watkins, Richard and Olive, Keith A.",
    editor = "Srednicki, M. A.",
    title = "{Calculations of Relic Densities in the Early Universe}",
    reportNumber = "UMN-TH-646/88",
    doi = "10.1016/0550-3213(88)90099-5",
    journal = "Nucl. Phys. B",
    volume = "310",
    pages = "693",
    year = "1988"
}

@article{Cirelli:2024ssz,
    author = "Cirelli, Marco and Strumia, Alessandro and Zupan, Jure",
    title = "{Dark Matter}",
    eprint = "2406.01705",
    archivePrefix = "arXiv",
    primaryClass = "hep-ph",
    month = "6",
    year = "2024"
}

@article{Gondolo:1990dk,
    author = "Gondolo, Paolo and Gelmini, Graciela",
    title = "{Cosmic abundances of stable particles: Improved analysis}",
    reportNumber = "UCLA-90-TEP-68",
    doi = "10.1016/0550-3213(91)90438-4",
    journal = "Nucl. Phys. B",
    volume = "360",
    pages = "145--179",
    year = "1991"
}

@article{Lee:1977ua,
    author = "Lee, Benjamin W. and Weinberg, Steven",
    editor = "Srednicki, M. A.",
    title = "{Cosmological Lower Bound on Heavy Neutrino Masses}",
    reportNumber = "FERMILAB-PUB-77-041-T",
    doi = "10.1103/PhysRevLett.39.165",
    journal = "Phys. Rev. Lett.",
    volume = "39",
    pages = "165--168",
    year = "1977"
}

@article{Bernstein:1985th,
    author = "Bernstein, Jeremy and Brown, Lowell S. and Feinberg, Gerald",
    title = "{The Cosmological Heavy Neutrino Problem Revisited}",
    reportNumber = "CU-TP-316",
    doi = "10.1103/PhysRevD.32.3261",
    journal = "Phys. Rev. D",
    volume = "32",
    pages = "3261",
    year = "1985"
}

@article{Scherrer:1985zt,
    author = "Scherrer, Robert J. and Turner, Michael S.",
    title = "{On the Relic, Cosmic Abundance of Stable Weakly Interacting Massive Particles}",
    reportNumber = "FERMILAB-PUB-85-163-A, EFI-85-76-CHICAGO",
    doi = "10.1103/PhysRevD.33.1585",
    journal = "Phys. Rev. D",
    volume = "33",
    pages = "1585",
    year = "1986",
    note = "[Erratum: Phys.Rev.D 34, 3263 (1986)]"
}

@article{Belanger:2024bro,
    author = "B\'elanger, Genevieve and Chakraborti, Sreemanti and G\'enolini, Yoann and Salati, Pierre",
    title = "{GeV-scale dark matter with p-wave Breit-Wigner enhanced annihilation}",
    eprint = "2401.02513",
    archivePrefix = "arXiv",
    primaryClass = "hep-ph",
    reportNumber = "LAPTH-002/24, IPPP/23/85",
    doi = "10.1103/PhysRevD.110.023039",
    journal = "Phys. Rev. D",
    volume = "110",
    number = "2",
    pages = "023039",
    year = "2024"
}

@article{Slatyer:2015jla,
    author = "Slatyer, Tracy R.",
    title = "{Indirect dark matter signatures in the cosmic dark ages. I. Generalizing the bound on s-wave dark matter annihilation from Planck results}",
    eprint = "1506.03811",
    archivePrefix = "arXiv",
    primaryClass = "hep-ph",
    reportNumber = "MIT-CTP-4682",
    doi = "10.1103/PhysRevD.93.023527",
    journal = "Phys. Rev. D",
    volume = "93",
    number = "2",
    pages = "023527",
    year = "2016"
}

@article{Binder:2017rgn,
    author = "Binder, Tobias and Bringmann, Torsten and Gustafsson, Michael and Hryczuk, Andrzej",
    title = "{Early kinetic decoupling of dark matter: when the standard way of calculating the thermal relic density fails}",
    eprint = "1706.07433",
    archivePrefix = "arXiv",
    primaryClass = "astro-ph.CO",
    doi = "10.1103/PhysRevD.96.115010",
    journal = "Phys. Rev. D",
    volume = "96",
    number = "11",
    pages = "115010",
    year = "2017",
    note = "[Erratum: Phys.Rev.D 101, 099901 (2020)]"
}

@article{Binder:2021bmg,
    author = "Binder, Tobias and Bringmann, Torsten and Gustafsson, Michael and Hryczuk, Andrzej",
    title = "{Dark matter relic abundance beyond kinetic equilibrium}",
    eprint = "2103.01944",
    archivePrefix = "arXiv",
    primaryClass = "hep-ph",
    doi = "10.1140/epjc/s10052-021-09357-5",
    journal = "Eur. Phys. J. C",
    volume = "81",
    pages = "577",
    year = "2021"
}

@article{Planck:2018vyg,
    author = "Aghanim, N. and others",
    collaboration = "Planck",
    title = "{Planck 2018 results. VI. Cosmological parameters}",
    eprint = "1807.06209",
    archivePrefix = "arXiv",
    primaryClass = "astro-ph.CO",
    doi = "10.1051/0004-6361/201833910",
    journal = "Astron. Astrophys.",
    volume = "641",
    pages = "A6",
    year = "2020",
    note = "[Erratum: Astron.Astrophys. 652, C4 (2021)]"
}

@article{Slatyer:2009yq,
    author = "Slatyer, Tracy R. and Padmanabhan, Nikhil and Finkbeiner, Douglas P.",
    title = "{CMB Constraints on WIMP Annihilation: Energy Absorption During the Recombination Epoch}",
    eprint = "0906.1197",
    archivePrefix = "arXiv",
    primaryClass = "astro-ph.CO",
    doi = "10.1103/PhysRevD.80.043526",
    journal = "Phys. Rev. D",
    volume = "80",
    pages = "043526",
    year = "2009"
}

@article{Cirelli:2020bpc,
    author = "Cirelli, Marco and Fornengo, Nicolao and Kavanagh, Bradley J. and Pinetti, Elena",
    title = "{Integral X-ray constraints on sub-GeV Dark Matter}",
    eprint = "2007.11493",
    archivePrefix = "arXiv",
    primaryClass = "hep-ph",
    doi = "10.1103/PhysRevD.103.063022",
    journal = "Phys. Rev. D",
    volume = "103",
    number = "6",
    pages = "063022",
    year = "2021"
}

@article{Essig:2013goa,
    author = "Essig, Rouven and Kuflik, Eric and McDermott, Samuel D. and Volansky, Tomer and Zurek, Kathryn M.",
    title = "{Constraining Light Dark Matter with Diffuse X-Ray and Gamma-Ray Observations}",
    eprint = "1309.4091",
    archivePrefix = "arXiv",
    primaryClass = "hep-ph",
    reportNumber = "YITP-SB-29-13, FERMILAB-PUB-13-377-A-T, MCTP-13-27",
    doi = "10.1007/JHEP11(2013)193",
    journal = "JHEP",
    volume = "11",
    pages = "193",
    year = "2013"
}

@article{Coogan_2020,
   title={Hazma: a python toolkit for studying indirect detection of sub-GeV dark matter},
   volume={2020},
   ISSN={1475-7516},
   url={http://dx.doi.org/10.1088/1475-7516/2020/01/056},
   DOI={10.1088/1475-7516/2020/01/056},
   number={01},
   journal={Journal of Cosmology and Astroparticle Physics},
   publisher={IOP Publishing},
   author={Coogan, Adam and Morrison, Logan and Profumo, Stefano},
   year={2020},
   month=jan, pages={056–056} }

@article{Gonzalez-Morales:2017jkx,
    author = "Gonzalez-Morales, Alma X. and Profumo, Stefano and Reynoso-C\'ordova, Javier",
    title = "{Prospects for indirect MeV Dark Matter detection with Gamma Rays in light of Cosmic Microwave Background Constraints}",
    eprint = "1705.00777",
    archivePrefix = "arXiv",
    primaryClass = "astro-ph.CO",
    doi = "10.1103/PhysRevD.96.063520",
    journal = "Phys. Rev. D",
    volume = "96",
    number = "6",
    pages = "063520",
    year = "2017"
}

@article{Alguero:2023zol,
    author = "Alguero, G. and Belanger, G. and Boudjema, F. and Chakraborti, S. and Goudelis, A. and Kraml, S. and Mjallal, A. and Pukhov, A.",
    title = "{micrOMEGAs 6.0: N-component dark matter}",
    eprint = "2312.14894",
    archivePrefix = "arXiv",
    primaryClass = "hep-ph",
    doi = "10.1016/j.cpc.2024.109133",
    journal = "Comput. Phys. Commun.",
    volume = "299",
    pages = "109133",
    year = "2024"
}

@article{Coogan:2022cdd,
    author = "Coogan, Adam and Morrison, Logan and Plehn, Tilman and Profumo, Stefano and Reimitz, Peter",
    title = "{Hazma meets HERWIG4DM: precision gamma-ray, neutrino, and positron spectra for light dark matter}",
    eprint = "2207.07634",
    archivePrefix = "arXiv",
    primaryClass = "hep-ph",
    doi = "10.1088/1475-7516/2022/11/033",
    journal = "JCAP",
    volume = "11",
    pages = "033",
    year = "2022"
}

@article{Wadekar:2021qae,
    author = "Wadekar, Digvijay and Wang, Zihui",
    title = "{Strong constraints on decay and annihilation of dark matter from heating of gas-rich dwarf galaxies}",
    eprint = "2111.08025",
    archivePrefix = "arXiv",
    primaryClass = "hep-ph",
    doi = "10.1103/PhysRevD.106.075007",
    journal = "Phys. Rev. D",
    volume = "106",
    number = "7",
    pages = "075007",
    year = "2022"
}

@article{Arkani-Hamed:1998jmv,
    author = "Arkani-Hamed, Nima and Dimopoulos, Savas and Dvali, G. R.",
    title = "{The Hierarchy problem and new dimensions at a millimeter}",
    eprint = "hep-ph/9803315",
    archivePrefix = "arXiv",
    reportNumber = "SLAC-PUB-7769, SU-ITP-98-13",
    doi = "10.1016/S0370-2693(98)00466-3",
    journal = "Phys. Lett. B",
    volume = "429",
    pages = "263--272",
    year = "1998"
}

@article{Chu:2022xuh,
    author = "Chu, Xiaoyong and Kuo, Jui-Lin and Pradler, Josef",
    title = "{Toward a full description of MeV dark matter decoupling: A self-consistent determination of relic abundance and Neff}",
    eprint = "2205.05714",
    archivePrefix = "arXiv",
    primaryClass = "hep-ph",
    doi = "10.1103/PhysRevD.106.055022",
    journal = "Phys. Rev. D",
    volume = "106",
    number = "5",
    pages = "055022",
    year = "2022"
}

@article{Sabti:2021reh,
    author = "Sabti, Nashwan and Alvey, James and Escudero, Miguel and Fairbairn, Malcolm and Blas, Diego",
    title = "{Addendum: Refined bounds on MeV-scale thermal dark sectors from BBN and the CMB}",
    eprint = "2107.11232",
    archivePrefix = "arXiv",
    primaryClass = "hep-ph",
    doi = "10.1088/1475-7516/2021/08/A01",
    journal = "JCAP",
    volume = "08",
    pages = "A01",
    year = "2021"
}

@article{Freese:1987wu,
    author = "Freese, Katherine and Frieman, Joshua A. and Gould, Andrew",
    title = "{Signal Modulation in Cold Dark Matter Detection}",
    reportNumber = "SLAC-PUB-4427, NSF-ITP-87-123",
    doi = "10.1103/PhysRevD.37.3388",
    journal = "Phys. Rev. D",
    volume = "37",
    pages = "3388--3405",
    year = "1988"
}

@article{Ryan-Weber:2007guk,
    author = "Ryan-Weber, Emma V. and Begum, Ayesha and Oosterloo, Tom and Pal, Sabyasachi and Irwin, Michael J. and Belokurov, Vasily and Evans, N. Wyn and Zucker, Daniel B.",
    title = "{The Local Group dwarf Leo T: HI on the brink of star formation}",
    eprint = "0711.2979",
    archivePrefix = "arXiv",
    primaryClass = "astro-ph",
    doi = "10.1111/j.1365-2966.2007.12734.x",
    journal = "Mon. Not. Roy. Astron. Soc.",
    volume = "384",
    pages = "53",
    year = "2008"
}

@article{Cirelli:2023tnx,
    author = "Cirelli, Marco and Fornengo, Nicolao and Koechler, Jordan and Pinetti, Elena and Roach, Brandon M.",
    title = "{Putting all the X in one basket: Updated X-ray constraints on sub-GeV Dark Matter}",
    eprint = "2303.08854",
    archivePrefix = "arXiv",
    primaryClass = "hep-ph",
    reportNumber = "FERMILAB-PUB-23-108-T",
    doi = "10.1088/1475-7516/2023/07/026",
    journal = "JCAP",
    volume = "07",
    pages = "026",
    year = "2023"
}

@article{Sabti:2019mhn,
    author = "Sabti, Nashwan and Alvey, James and Escudero, Miguel and Fairbairn, Malcolm and Blas, Diego",
    title = "{Refined Bounds on MeV-scale Thermal Dark Sectors from BBN and the CMB}",
    eprint = "1910.01649",
    archivePrefix = "arXiv",
    primaryClass = "hep-ph",
    reportNumber = "KCL-2019-75",
    doi = "10.1088/1475-7516/2020/01/004",
    journal = "JCAP",
    volume = "01",
    pages = "004",
    year = "2020"
}

@article{Arkani-Hamed:1998sfv,
    author = "Arkani-Hamed, Nima and Dimopoulos, Savas and Dvali, G. R.",
    title = "{Phenomenology, astrophysics and cosmology of theories with submillimeter dimensions and TeV scale quantum gravity}",
    eprint = "hep-ph/9807344",
    archivePrefix = "arXiv",
    reportNumber = "SLAC-PUB-7864, SU-ITP-98-142, IC-98-44",
    doi = "10.1103/PhysRevD.59.086004",
    journal = "Phys. Rev. D",
    volume = "59",
    pages = "086004",
    year = "1999"
}

@article{Kakizaki:2005en,
    author = "Kakizaki, Mitsuru and Matsumoto, Shigeki and Sato, Yoshio and Senami, Masato",
    title = "{Significant effects of second KK particles on LKP dark matter physics}",
    eprint = "hep-ph/0502059",
    archivePrefix = "arXiv",
    reportNumber = "ICRR-514-2004-12, STUPP-05-178",
    doi = "10.1103/PhysRevD.71.123522",
    journal = "Phys. Rev. D",
    volume = "71",
    pages = "123522",
    year = "2005"
}

@article{Zhao_2016,
   title={Constraint on the velocity dependent dark matter annihilation cross section from Fermi-LAT observations of dwarf galaxies},
   volume={93},
   ISSN={2470-0029},
   url={http://dx.doi.org/10.1103/PhysRevD.93.083513},
   DOI={10.1103/physrevd.93.083513},
   number={8},
   journal={Physical Review D},
   publisher={American Physical Society (APS)},
   author={Zhao, Yi and Bi, Xiao-Jun and Jia, Huan-Yu and Yin, Peng-Fei and Zhu, Feng-Rong},
   year={2016},
   month=apr }

@article{Fermi-LAT:2015att,
    author = "Ackermann, M. and others",
    collaboration = "Fermi-LAT",
    title = "{Searching for Dark Matter Annihilation from Milky Way Dwarf Spheroidal Galaxies with Six Years of Fermi Large Area Telescope Data}",
    eprint = "1503.02641",
    archivePrefix = "arXiv",
    primaryClass = "astro-ph.HE",
    reportNumber = "FERMILAB-PUB-15-081-AE",
    doi = "10.1103/PhysRevLett.115.231301",
    journal = "Phys. Rev. Lett.",
    volume = "115",
    number = "23",
    pages = "231301",
    year = "2015"
}

@article{Lavis:2023jju,
    author = "Lavis, Natasha and Sarkis, Michael and Beck, Geoff and Knowles, Kenda",
    title = "{Radio-frequency WIMP search with the MeerKAT galaxy cluster legacy survey}",
    eprint = "2308.08351",
    archivePrefix = "arXiv",
    primaryClass = "astro-ph.CO",
    doi = "10.1103/PhysRevD.108.123536",
    journal = "Phys. Rev. D",
    volume = "108",
    number = "12",
    pages = "123536",
    year = "2023"
}

@article{Sarkis:2024zjg,
    author = "Sarkis, Michael and Beck, Geoff",
    title = "{DarkMatters: A powerful tool for WIMPy analysis}",
    eprint = "2408.07053",
    archivePrefix = "arXiv",
    primaryClass = "hep-ph",
    doi = "10.1016/j.dark.2024.101745",
    journal = "Phys. Dark Univ.",
    volume = "47",
    pages = "101745",
    year = "2025"
}

@article{Feldman:2008xs,
    author = "Feldman, Daniel and Liu, Zuowei and Nath, Pran",
    title = "{PAMELA Positron Excess as a Signal from the Hidden Sector}",
    eprint = "0810.5762",
    archivePrefix = "arXiv",
    primaryClass = "hep-ph",
    doi = "10.1103/PhysRevD.79.063509",
    journal = "Phys. Rev. D",
    volume = "79",
    pages = "063509",
    year = "2009"
}

@article{Guo:2009aj,
    author = "Guo, Wan-Lei and Wu, Yue-Liang",
    title = "{Enhancement of Dark Matter Annihilation via Breit-Wigner Resonance}",
    eprint = "0901.1450",
    archivePrefix = "arXiv",
    primaryClass = "hep-ph",
    doi = "10.1103/PhysRevD.79.055012",
    journal = "Phys. Rev. D",
    volume = "79",
    pages = "055012",
    year = "2009"
}

@article{AlbornozVasquez:2011js,
    author = "Albornoz Vasquez, Daniel and Belanger, Genevieve and Boehm, Celine",
    title = "{Astrophysical limits on light NMSSM neutralinos}",
    eprint = "1107.1614",
    archivePrefix = "arXiv",
    primaryClass = "hep-ph",
    reportNumber = "LAPTH-025-11, IPPP-11-37, DCPT-11-74",
    doi = "10.1103/PhysRevD.84.095008",
    journal = "Phys. Rev. D",
    volume = "84",
    pages = "095008",
    year = "2011"
}

@article{Duch:2017nbe,
    author = "Duch, Mateusz and Grzadkowski, Bohdan",
    title = "{Resonance enhancement of dark matter interactions: the case for early kinetic decoupling and velocity dependent resonance width}",
    eprint = "1705.10777",
    archivePrefix = "arXiv",
    primaryClass = "hep-ph",
    doi = "10.1007/JHEP09(2017)159",
    journal = "JHEP",
    volume = "09",
    pages = "159",
    year = "2017"
}

@article{Binder:2022pmf,
    author = "Binder, Tobias and Chakraborti, Sreemanti and Matsumoto, Shigeki and Watanabe, Yu",
    title = "{A global analysis of resonance-enhanced light scalar dark matter}",
    eprint = "2205.10149",
    archivePrefix = "arXiv",
    primaryClass = "hep-ph",
    reportNumber = "LAPTH-029/22",
    doi = "10.1007/JHEP01(2023)106",
    journal = "JHEP",
    volume = "01",
    pages = "106",
    year = "2023"
}

@article{Bernreuther:2020koj,
    author = "Bernreuther, Elias and Heeba, Saniya and Kahlhoefer, Felix",
    title = "{Resonant sub-GeV Dirac dark matter}",
    eprint = "2010.14522",
    archivePrefix = "arXiv",
    primaryClass = "hep-ph",
    reportNumber = "P3H-20-062, TTK-20-38",
    doi = "10.1088/1475-7516/2021/03/040",
    journal = "JCAP",
    volume = "03",
    pages = "040",
    year = "2021"
}

@article{Ding:2021sbj,
    author = "Ding, Yu-Chen and Ku, Yu-Lin and Wei, Chun-Cheng and Zhou, Yu-Feng",
    title = "{Consistent explanation for the cosmic-ray positron excess in p-wave Breit\textendash{}Wigner enhanced dark matter annihilation}",
    eprint = "2110.10388",
    archivePrefix = "arXiv",
    primaryClass = "hep-ph",
    doi = "10.1140/epjc/s10052-022-10048-y",
    journal = "Eur. Phys. J. C",
    volume = "82",
    number = "2",
    pages = "126",
    year = "2022"
}

@article{Chen:2024njd,
    author = "Chen, Yu-Tong and Matsumoto, Shigeki and Tang, Tian-Peng and Tsai, Yue-Lin Sming and Wu, Lei",
    title = "{Light thermal dark matter beyond p-wave annihilation in minimal Higgs portal model}",
    eprint = "2403.02721",
    archivePrefix = "arXiv",
    primaryClass = "hep-ph",
    doi = "10.1007/JHEP05(2024)281",
    journal = "JHEP",
    volume = "05",
    pages = "281",
    year = "2024"
}

@article{Balan:2024cmq,
    author = "Balan, Sowmiya and others",
    title = "{Resonant or asymmetric: the status of sub-GeV dark matter}",
    eprint = "2405.17548",
    archivePrefix = "arXiv",
    primaryClass = "hep-ph",
    reportNumber = "TTP24-015, P3H-24-033",
    doi = "10.1088/1475-7516/2025/01/053",
    journal = "JCAP",
    volume = "01",
    pages = "053",
    year = "2025"
}

@article{Adams_2018,
       author = {{Adams}, Elizabeth A.~K. and {Oosterloo}, Tom A.},
        title = "{Deep neutral hydrogen observations of Leo T with the Westerbork Synthesis Radio Telescope}",
      journal = {\aap},
         year = 2018,
        month = apr,
       volume = {612},
          eid = {A26},
        pages = {A26},
          doi = {10.1051/0004-6361/201732017},
archivePrefix = {arXiv},
       eprint = {1712.06636},
 primaryClass = {astro-ph.GA},
       adsurl = {https://ui.adsabs.harvard.edu/abs/2018A&A...612A..26A}
}

@article{Knowles:2021nvz,
    author = "Knowles, K. and others",
    title = "{The MeerKAT Galaxy Cluster Legacy Survey - I. Survey Overview and Highlights}",
    eprint = "2111.05673",
    archivePrefix = "arXiv",
    primaryClass = "astro-ph.GA",
    doi = "10.1051/0004-6361/202141488",
    journal = "Astron. Astrophys.",
    volume = "657",
    pages = "A56",
    year = "2022"
}

@inproceedings{Jonas:2018Jr,
  author = "Jonas, Justin",
  title = "{The MeerKAT Radio Telescope}",
  doi = "10.22323/1.277.0001",
  booktitle = "Proceedings of MeerKAT Science: On the Pathway to the SKA {\textemdash} PoS(MeerKAT2016)",
  year = 2018,
  volume = "277",
  pages = "001"
}

@inproceedings{Beck:2023oza,
    author = "Beck, Geoff and Makhathini, Sphesihle",
    title = "{Just a MeerKAT, or a dark matter machine?}",
    eprint = "2301.07910",
    archivePrefix = "arXiv",
    primaryClass = "hep-ph",
    month = "1",
    year = "2023"
}

\end{document}